\documentclass{emulateapj}

\usepackage{natbib}
\newcommand{\unit}[1]{\mathrm{#1}}

\newcommand{\wpp}{w_{\mathrm{p}}}
\newcommand{\rp}{(r_{\mathrm{p}})}

\newcommand{\rhor}{\rho(\textbf{r})}
\newcommand{\ngal}{\bar{n}_{\mathrm{g}}}
\newcommand{\Nsat}{N_\mathrm{sat}}

\newcommand{\arcs}{\unit{arcsec}}

\newcommand{\hmpc}{h^{-1}\mathrm{Mpc}}
\newcommand{\hmpcVol}{h^{3}\mathrm{Mpc}^{-3}}
\newcommand{\hkpc}{h^{-1}\mathrm{kpc}}
\newcommand{\hMsun}{h^{-1}M_{\odot}}

\newcommand{\Mmin}{M_\mathrm{min}}
\newcommand{\Mzero}{M_\mathrm{0}}
\newcommand{\Mone}{M_\mathrm{1}}
\newcommand{\Mstar}{M_\ast}
\newcommand{\Cgal}{c_\mathrm{gal}}
\newcommand{\fgal}{f_\mathrm{gal}}
\newcommand{\onehalo}{\mathrm{1halo}}
\newcommand{\twohalo}{\mathrm{2halo}}
\newcommand{\expon}{\mathrm{exp}}

\newcommand{\Rvir}{R_\mathrm{vir}}
\usepackage{epsfig}
\usepackage{graphicx} 
\usepackage{rotating}

\begin{document}

\title{MODELING THE VERY SMALL-SCALE CLUSTERING OF LUMINOUS RED GALAXIES}
\author{
	Douglas~F.~Watson\altaffilmark{1},
	Andreas~A.~Berlind\altaffilmark{1},
	Cameron~K.~McBride\altaffilmark{1},
	Morad~Masjedi\altaffilmark{2}
}

\altaffiltext{1}{Department of Physics and Astronomy, Vanderbilt University, 1807 Station B, Nashville, TN 37235}
\altaffiltext{2}{Goldman Sachs \& Co., 85 Broad Street, New York, NY 10004}

\begin{abstract}\label{abstract}
We model the small-scale clustering of luminous red galaxies (LRGs; \citealt{masjedi06a}) in the Sloan Digital Sky Survey (SDSS).  Specifically, we use the halo occupation distribution (HOD) formalism to model the projected two-point correlation function of LRGs on scales well within the sizes of their host halos ($0.016 \hmpc \leq \emph{r} \leq 0.42 \hmpc$).  We start by varying $P(N|M)$, the probability distribution that a dark matter halo of mass $M$ contains $N$ LRGs, and assuming that the radial distribution of satellite LRGs within halos traces the NFW dark matter density profile.  We find that varying $P(N|M)$ alone is not sufficient to match the small-scale data.  We next allow the concentration of satellite LRG galaxies to differ from that of dark matter and find that this is also not sufficient.  Finally, we relax the assumption of an NFW profile and allow the inner slope of the density profile to vary.  We find that this model provides a good fit to the data and the resulting value of the slope is $-2.17\pm 0.12$.  The radial density profile of satellite LRGs within halos is thus not compatible with that of the underlying dark matter, but rather is closer to an isothermal distribution.
\end{abstract}

\keywords{cosmology: theory --- galaxies: elliptical and lenticular, cD --- galaxies: fundamental parameters --- large-scale structure of universe --- methods: statistical --- surveys}


\section{INTRODUCTION}\label{intro}

Galaxy clustering provides important insight into the physics of galaxy formation, and can be used to probe a broad range of physical phenomena on different cosmological scales.  On scales smaller than the virial radii of the largest halos ($\Rvir\lesssim 1$ Mpc) complex dynamical processes are occurring, such as dynamical friction, tidal forces, and galaxy mergers.  These most certainly have an effect on the way galaxies cluster, and will give rise to features in the correlation function $\xi(r)$ on small scales.

The Sloan Digital Sky Survey \citep{york00a} has produced the largest ever spectroscopic sample of luminous red galaxies (LRGs).  The clustering of these galaxies has been considered in great detail on both intermediate  \citep{zehavi05b} and very large scales \citep{eisenstein05b}.  However, it is especially difficult to measure the correlation function on small scales due to ``fiber collision'' incompleteness, and deblending issues.  Masjedi et al. (2006, hereinafter M06) were able to overcome these observational impasses.  They corrected for fiber collisions by cross-correlating the spectroscopic sample with the imaging sample and testing the results against mock data sets.  They addressed deblending errors by introducing artificial galaxies of known magnitudes into the data and studying how they are recovered by the SDSS software.  These adjustments allowed them to properly measure $\xi(r)$ down to a separation of \emph{r} $\sim 15\hkpc$.  The clustering on these scales has also been measured for lower luminosity galaxies \citep{wang06, li09}, though these studies did not include deblending corrections, which M06 showed can be significant.

In this paper, we model this recently measured very small-scale LRG clustering.  On intermediate scales ($0.3-40 \hmpc$), \citet{zehavi05b} measured the two-point correlation function for 35,000 LRGs.  \citet{zheng08} modeled this data using the halo occupation distribution (HOD; see, e.g., \citealt{peacock00a, scoccimarro01a, berlind02a, cooray02}) framework and found a nice fit.  However, the extrapolation of their best-fit model to smaller separations does not agree with the M06 small-scale data: it predicts a correlation function that is too low (see M06, Fig.~4).  Our motivation is to model these innermost data points ($0.016-0.42h^{-1}$ Mpc) to see if we can find a model that works.  The paper is laid out as follows.  In \S~\ref{data}, we review the M06 measurement.  In \S~\ref{method}, we discuss our method for modeling the small-scale correlation function.  In \S~\ref{results}, we discuss our results in a sequential format: in \S~\ref{varyHOD}, we use four free parameters from the probability distribution, $P(N|M)$; in \S~\ref{varyCONC}, we introduce the concentration of satellite LRGs as a new free parameter; in \S~\ref{varyPROFILE}, we allow the inner slope of the density profile of satellite LRGs to vary.  Finally, in \S~\ref{discussion}, we discuss the implications of our results.


\section{DATA}\label{data}

M06 measured the small-scale ($0.016-8\hmpc$) projected two-point correlation function for a volume-limited sample of 24,520 luminous red galaxies in the SDSS.  The luminosity range of LRGs in the sample was $-23.2<M_g<-21.2$ and the redshift range was $0.16<z<0.36$.  Measuring the correlation function on such small scales is non-trivial, and requires overcoming two main observational hurdles: fiber collisions and deblending.

The SDSS spectroscopic sample is incomplete due to the physical size of the fiber-optic cables used to take spectra of targeted galaxies.  If two galaxies are closer than 55 $\arcs$ on the sky, they cannot both get measured spectra.  These ``fiber collisions'' thus result in a minimum possible pair separation of 55 $\arcs$.  Fiber collision incompleteness is reduced by the SDSS tiling method, which overlaps the spectroscopic plates to achieve continuous sky coverage.  However, this still results in $\sim 7 \%$ of targeted galaxies without measured redshifts.  Naturally, this incompleteness affects galaxy clustering measurements most severely at very small scales.  To get around this problem, M06 cross-correlated the spectroscopic LRG sample with the entire sample of LRG targets in the SDSS imaging.  For every LRG from the spectroscopic sample, nearby LRG targets (whether or not they have an observed spectrum) were considered to be at the same redshift as the spectroscopically observed LRG.  This allowed M06 to assign absolute magnitudes to the LRG targets and thus decide if they made it into the volume-limited sample.  M06 then statistically removed the contribution of galaxies that were not at the same redshift, but were considered to be by the algorithm, by constructing random samples with the same redshift distribution as the spectroscopic sample, and cross-correlating them with the LRG imaging sample.  M06 tested this procedure on mock galaxy catalogs and found that it successfully recovers the LRG auto-correlation function.

The SDSS photometry of LRG galaxies is biased in cases where pairs of LRGs are separated by tens of kpc or less.  These galaxies have a region of overlap, and the light contained in this region needs to be properly distributed between the LRG pair.  This process is called deblending.  Since the LRG sample is defined by luminosity cuts ($-23.2<M_g<-21.2$), any systematic errors in deblending will lead to incorrect measurements of the correlation function.  M06 tested the deblending method by introducing artificial, overlapping galaxies of assigned magnitudes into the SDSS imaging.  They then deblended the images using the SDSS imaging pipeline $PHOTO$ \citep{lupton01}, and discovered that too much light was being systematically allocated to the fainter galaxy of the pair.  In many cases, this pushed the fainter galaxy above the luminosity cut and into the LRG sample, whereas it should have stayed out.  This photometric bias thus led to a boost in the correlation function on small scales.  M06 corrected their correlation function for deblending errors using the results of their tests with artificial data.

After applying these corrections and deprojecting $\wpp\rp$, M06 found that the LRG correlation function on small scales is a continuation of the $\xi(r)\propto r^{-2}$ power law found previously for larger scales \citep{zehavi05b}.  Only their smallest scale data point ($\sim 10\hkpc$) shows a downturn, which presumably occurs because at such a small scale, it is no longer possible to distinguish two merging LRGs from each other.  In this paper, we use M06's measurements of the projected correlation function $\wpp\rp$ from $16$ to $420\hkpc$, which consist of eight data points.  Restricting ourselves to this range guarantees that we are safely within the 1-halo regime (i.e., all LRG pairs are coming from within the same halos).  We use the full covariance matrix for this data, which M06 estimated using jackknife resampling.

We incorporate the measured number density of galaxies as an additional constraint in our modeling.  The number density of LRGs with $M_{g} < -21.2$ is $\ngal=9.73\times10^{-5} \hmpcVol$ \citep{zheng08} and this provides us with a ninth data point.  We estimate an error for this number density using 50 jackknife samples on the sky and obtain $\sigma_{\ngal}= 1.46\times10^{-6}\hmpcVol$.


\section{METHOD}\label{method}

\subsection{The Halo Occupation Distribution}\label{HOD}

The HOD framework characterizes the bias between galaxies (of any class) and mass and is completely defined by (1) the probability distribution $P(N|M)$ that a virialized  dark matter halo of mass $M$ will host $N$ galaxies, (2) the relative spatial distribution of the galaxies and dark matter within their host halo, and (3) the relative velocity distribution between the galaxies and dark matter within the halo \citep{berlind02a}.  We may neglect (3) in this study due to the fact that $\wpp\rp$ is velocity independent.  

The first moment of $P(N|M)$ is the mean number of galaxies as a function of halo mass $\langle N \rangle_{M}$, and we parametrize it as a sum of a central and a satellite component \citep{kravtsov04b, tinker07, zheng08}.  We assume that there is a minimum halo mass cut-off (hereinafter referred to as $\Mmin$) below which a halo will always be empty and above which a halo will always contain at least a single central galaxy.  We next assume that the mean number of satellite galaxies is a power-law function of mass with a low-mass exponential cutoff at $\Mzero$.  The power-law slope is $\alpha$ and the normalization is $\Mone$, which represents the mass where halos contain, on average, a single satellite galaxy.  Specifically, the average number of galaxies as a function of mass is $\langle N \rangle _{M} = 1 + \langle \Nsat \rangle _{M}$, where
\begin{equation}
\langle \Nsat \rangle _{M} = \expon[-{\Mzero}/(M_{\mathrm{halo}}-\Mmin)]\times(M_{\mathrm{halo}}/\Mone)^{\alpha}
\end{equation} 

The correlation function $\xi(r)$ depends on the second moment of $P(N|M)$ and so specifying $\langle N \rangle_{M}$ is not sufficient.  We assume that the actual number of satellite galaxies in a halo of mass $M$ follows a Poisson distribution of mean $\langle \Nsat \rangle$.  Therefore, the second moment is $\langle \Nsat(\Nsat-1) \rangle _{M} = \langle \Nsat \rangle ^{2}_{M}$.  Our choice of a Poisson distribution is motivated by theoretical results \citep{kravtsov04b,zheng05}.  We note, however, that our results will not be very sensitive to the specific form of the satellite galaxy distribution because of the high mass regime that we are considering in this study.  LRGs live in halos of mass greater than the non-linear mass $\Mstar$ \citep{zheng08}, which means that they are on the exponentially declining part of the halo mass function.  As a result, most of our LRG pairs will be central-satellite pairs in halos of mass $\sim\Mone$, rather than satellite-satellite pairs in higher mass halos.

In addition to $P(N|M)$, we must also characterize the spatial distribution of galaxies within halos.  We naturally assume that the central galaxy sits at the center of its halo.  As for satellite galaxies, we assume at first that they trace the density distribution of dark matter within their halo, but we eventually relax that assumption, as we describe in \S~\ref{results}.  We assume that halos have dark matter density profiles that are described by the NFW relation $\rhor\propto(r/r_s)^{-1}(1+r/r_s)^{-2}$ \citep{nfw97}.  The NFW scale radius $r_s$ controls where the profile transitions from $r^{-1}$ in the inner parts to $r^{-3}$ in the outer parts, but we parametrize it instead through the concentration parameter, which is defined as the ratio of a halo's virial radius to its scale radius ($c\equiv \Rvir/r_s$).  Finally, we adopt the concentration - mass relation given by \citet{zheng07} for the modification of \citet{bullock01}: $c=\frac{c_{0}}{(1 + z)}\times(\frac{M_\mathrm{halo}}{M_{\ast}})^{-\beta}$, where $c_{0}=11$, $M_{\ast}$ is the non-linear mass at the median redshift ($\Mstar = 1.2\times 10^{12} \hMsun, z = 0.286$) of the sample for our choice of cosmology, and $\beta = .13$. 

\subsection{The Galaxy Number Density}\label{ngal}

We can calculate the galaxy number density for a given HOD by integrating over halo mass and weighting the abundance of halos by the mean occupation of galaxies $\langle N \rangle _{M}$:
\begin{equation}
\ngal = \int_{\Mmin}^\infty dM\frac{dn}{dM}\langle N \rangle _{M},
\end{equation}
where $dn/dM$ is the differential halo mass function.  We adopt the \citet{warren06} halo mass function with the following cosmological model: $\Omega_{m}=0.25, \Omega_{\Lambda}=0.75, \Omega_{b}=0.04, h_{0}=0.7, \sigma_{8}=0.8, n_{s}=1.0$.

\subsection{The Galaxy 2-point Correlation Function}\label{2pt_xigg}

In the halo model, the galaxy two-point correlation function is given as the sum of the ``1-halo'' and ``2-halo'' terms \citep{zheng04a},
\begin{equation}
\xi_{\mathrm{gg}}(r)=\xi_{\mathrm{gg}}^{\onehalo}+\xi_{\mathrm{gg}}^{\twohalo} + 1 
\end{equation}
The 2-halo term is the contribution of galaxy pairs found in separate dark matter halos.  Therefore, at scales smaller than the virial diameter of the smallest halos considered, the 1-halo term completely dominates the correlation function.  \citet{zheng08} found that the minimum halo mass for LRGs is $\Mmin\cong10^{13.7}M_{\odot}$, which corresponds to virial radii of $0.8-0.9\hmpc$.  Therefore, it is sufficient to only consider the 1-halo term when modeling $\xi(r)$ for our small-scale data.

The 1-halo term of the two-point correlation function can be written as an integral over halo mass, where at each mass we add the contribution of central-satellite and satellite-satellite pairs:
\begin{eqnarray}
1 + \xi_{\mathrm{gg}}^{\onehalo}(r) = \frac{1}{2\pi r^{2}\ngal^2}\int_{\Mmin}^\infty dM\frac{dn}{dM} \\
\times \Big[\langle \Nsat \rangle _{M} F_{\mathrm{cs}}(r)+\frac{\langle \Nsat(\Nsat-1) \rangle _{M}}{2} F_{\mathrm{ss}}(r) \Big],\nonumber
\end{eqnarray}
where $F_{\mathrm{cs}}(r)$ and $F_{\mathrm{ss}}(r)$ are the pair separation distributions of central-satellite and satellite-satellite pairs, respectively.  The pair distribution for central-satellite pairs is essentially the same as the density profile of satellite  galaxies, whereas the satellite-satellite pair distribution is equivalent to a convolution of the density profile with itself.  We use the \citet{sheth01} calculation for the convolution of a truncated NFW profile with itself.  As we discussed in \S~\ref{HOD}, central-satellite pairs will dominate the LRG correlation function due to the large halo masses involved.  Therefore, the shape of the small-scale LRG correlation function should be very close to the shape of the satellite galaxy density profile.

M06 measured the projected correlation function, so we must transform our theoretical $\xi(r)$ into $\wpp\rp$ by integrating along the line of sight \citep{davis83,zehavi04a}: 
\begin{equation}
\wpp\rp = 2\int_0^\infty \xi\Big(\sqrt{r_{\mathrm{p}}^2+y^2}\Big)dy .
\end{equation}
Since we need to integrate $\xi(r)$ to large radii in order to get $\wpp\rp$,  we cannot completely ignore the 2-halo term.  However, $\xi(r)$ is a rapidly declining function of $r$ and so $\wpp\rp$ at the small scales we are considering is not sensitive to variations in the 2-halo term.  For this reason, we use the best-fit 2-halo term from \citet{zheng08} instead of calculating it explicitly each time we vary our HOD parameters.  We tested this by shifting the amplitude of the 2-halo term by 20\% in each direction and we found that $\wpp\rp$ is not appreciably affected.

\subsection{Probing the Parameter Space}\label{MCMC}

Now that we have measurements of $\ngal$ and $\wpp\rp$, as well as a mechanism to predict these quantities from a given set of HOD parameters (e.g., $\Mmin$, $\Mzero$, $\Mone$, $\alpha$), we can probe the parameter space and find the region that gives a good fit to the data.  We use a Markov Chain Monte Carlo (MCMC) method for this purpose.  Specifically, we adopt the Metropolis-Hastings algorithm, which works as follows.  The free parameters are given initial values and $\chi^2$ is computed for this starting point in parameter space.  Steps are then chosen for the parameters and $\chi^2$ is computed for this new location.  The new location is added to the chain if $\chi^{2}_{new}<\chi^{2}_{old}$ or $n < \expon[-(\chi^{2}_{new}-\chi^{2}_{old})/2]$, where n is a random number between 0 and 1.  If these conditions are not met, then the old location is repeated in the chain.  This process then repeats until the chain has converged.  We test for convergence by starting three chains with different initial parameter values and checking whether their final parameter distribution functions are in agreement.  We assume uniform priors for all parameters and do not impose restrictions on their ranges.

Once a given chain has converged, we can plot the histogram of parameter values in the chain.  The most likely value for each parameter is given by the mean of its distribution, and the associated errors are given by the extrema of the middle $68.3 \%$ of the distribution.  The best-fit parameters correspond to the combination of parameter values for which $\chi^{2}$ is a minimum.  For more details on MCMC techniques, see \citet{dunkley05}.


\section{RESULTS}\label{results}

\subsection{Varying $P(N|M)$} \label{varyHOD}

\begin{figure}
\begin{center}
\includegraphics[scale=0.4,angle=90]{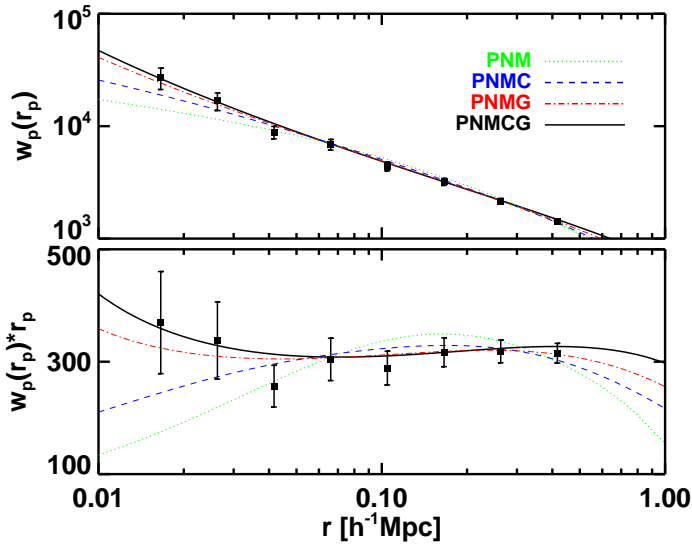}
\caption{Model fits to the projected correlation function of LRGs.  Points show the \citet{masjedi06a} $\wpp\rp$ measurements along with their jackknife errors.  The four curves show the best-fit models for four sets of free parameters.  The PNM model ({\it dotted green curve}) uses 4 free parameters that describe the probability distribution $P(N|M)$ - $\Mmin$, $\Mzero$, $\Mone$, and $\alpha$.  The PNMC model ({\it dashed blue curve}) replaces $\Mzero$ with $\fgal$, which is the concentration of the density profile of LRG satellites relative to dark matter.  The PNMG model ({\it dashed-dotted red curve}) replaces this with the inner slope of the density profile $\gamma$.  The PNMCG model ({\it solid black curve}) is the same as the PNMG model with the concentration of satellite LRGs $\fgal$ added as a fifth free parameter.  The $\emph{top panel}$ shows $\wpp\rp$, and the $\emph{bottom panel}$ shows the residuals from an $r_\mathrm{p}^{-1}$ power law.}
\label{fig:wpgg_Slope}
\end{center}
\end{figure}

As discussed in \S~\ref{intro}, the \citet{zheng08} fit to the intermediate-scale LRG correlation function does not match the M06 data points when extrapolated inwards.  We first investigate whether we can achieve a good fit to these new M06 small-scale data points while only varying the $P(N|M)$ part of the HOD.  We vary the following parameters (also defined in \S~\ref{HOD}):
\begin{enumerate}
\item $\Mmin$ - the minimum halo mass to contain a central galaxy.
\item $\Mzero$ - the minimum halo mass to contain satellite galaxies.
\item $\Mone$ - the halo mass to contain, on average, one satellite galaxy.
\item $\alpha$ - the slope of the power-law relation between the mean number of satellite galaxies and halo mass.
\end{enumerate}
We refer to this model as PNM.  Our MCMC rapidly converges and we find a best-fit model with a reduced $\chi^2$ of 6.06 ($\chi^2$ of 30.3 with 5 degrees of freedom).  The dotted green curves in Figure~\ref{fig:wpgg_Slope} show this best fit.  It is clear that the model provides a poor fit to the data, as it deviates downward from a power law on small scales.  We therefore find that by varying the $P(N|M)$ free parameters alone we are unable to reproduce the innermost M06 data points.  We have essentially reproduced the discrepancy between the very small-scale M06 data and the \citet{zheng08} modeling, and thus shown that including the small-scale points in the fit does not repair the discrepancy.

\subsection{Varying the Concentration of Satellite Galaxies}\label{varyCONC}

Since we cannot reproduce the small-scale LRG clustering by varying the $P(N|M)$ distribution, we naturally set our sights next on the radial distribution of LRG satellites within their dark matter halos.  As described in \S~\ref{HOD}, we have assumed that these galaxies trace the dark matter halo density distribution, which is in turn described by an NFW density profile.  The simplest change we can make is to allow galaxies to have a different NFW concentration than the halos they occupy.  We thus introduce a new free parameter $\fgal$ that relates the satellite galaxy concentration $\Cgal$ to that of the dark matter halo $c$:
\begin{equation}
\Cgal =\fgal\times c .
\end{equation}
In the previous 4 parameter PNM model we found that $\Mzero$ was very poorly constrained.  In order to keep the same number of free parameters, we fix $\Mzero$ to $\Mmin$, setting the exponential cut-off for satellite galaxies to occur at $\Mmin$.  Therefore, we now vary the following 4 free parameters: $\Mmin$, $\Mone$, $\alpha$, and $\fgal$.  We refer to this model as PNMC.

We find a best-fit model with a reduced $\chi^2$ of 2.52 ($\chi^2$ of 12.6 with 5 degrees of freedom).  The dashed blue curves in Figure~\ref{fig:wpgg_Slope} show this best fit.  The PNMC model clearly does better than the PNM model in explaining the small-scale LRG clustering; however, it still provides a poor fit.  We find that $\fgal$ values of $\sim 5-10$ are preferred, showing that LRGs are more concentrated than the dark matter for this PNMC model.  This makes sense because increasing $\Cgal$ means that we are adding more satellite galaxies towards the center of halos.  This forces the scale radius inwards and boosts the amplitude of the inner part of $\xi(r)$.  However, while simply moving more galaxies towards the center may aid in fitting the very inner most 2-3 data points, this can result in a poorer fit to the outermost data points.  In other words, varying $\Cgal$ can shift $\wpp\rp$ in the $r_\mathrm{p}$ direction, but it cannot alter its $\emph{shape}$, which is fundamentally not a power law in the case of an NFW satellite profile (see dashed blue curve in bottom panel of Fig.~\ref{fig:wpgg_Slope}).

\subsection{Varying the Density Profile}\label{varyPROFILE}

Adopting an NFW form for the density profile of satellite LRGs is not capable of reproducing the small-scale correlation function, no matter what concentration we use.  We therefore relax the NFW assumption by allowing the inner slope of the profile to vary.  Recall that the NFW profile has a logarithmic slope $d\ln \rho/d\ln r$ of -1 at scales much less than the scale radius $r_s$, and -3 at scales much larger than $r_s$.  We assume a new density profile for satellite LRG galaxies that is similar to NFW, except that the inner slope is no longer fixed to -1, but is a new free parameter $-\gamma$:
\begin{equation}\label{eq:profile_gamma}
\rho(r) = \frac{\rho_{s}}{\Big(\frac{r}{r_{s}} \Big)^{\gamma}  \Big(1+\frac{r}{r_{s}} \Big)^{3 - \gamma}} .
\end{equation}
This reduces to NFW for $\gamma=1$.  A model of this form has been used by papers that study the inner slope of the dark matter density profile (e.g., \citealt{fukushige04,reed05}).

In order to use this new profile in our modeling, we need to compute the pair distributions $F_{\mathrm{cs}}(r)$ and $F_{\mathrm{ss}}(r)$, as described in \S~\ref{2pt_xigg}.  While $F_{\mathrm{cs}}(r)$ is the profile itself, $F_{\mathrm{ss}}(r)$ is the convolution of the profile with itself and quite non-trivial to calculate analytically for arbitrary values of $\gamma$ and concentration.  We therefore calculate $F_{\mathrm{ss}}(r)$ in a numerical fashion.  We make a dense grid of $\gamma$ and concentration values, and at each grid point we create an artificial spherical halo by putting down 30k particles that satisfy the radial profile for that grid point.  We then measure the pair distribution by counting all the particle pairs in our constructed halo.  Once we have a table of $F_{\mathrm{ss}}(r)$ functions on our grid, we can estimate $F_{\mathrm{ss}}(r)$ for any values of $\gamma$ and concentration by interpolating in the grid.

As before, we wish to keep the same number of free parameters in order to more fairly compare different models.  We thus keep $\Mzero$ fixed to $\Mmin$ and we keep $\fgal$ fixed to unity.   Therefore, we now vary the following 4 free parameters: $\Mmin$, $\Mone$, $\alpha$, and $\gamma$, and we refer to this model as PNMG.  We find a best-fit model with a reduced $\chi^2$ of 0.82 ($\chi^2$ of 4.11 with 5 degrees of freedom).  The dashed-dotted red curves in Figure~\ref{fig:wpgg_Slope} show this best fit.  The PNMG model is clearly successful in fitting the M06 small-scale data.  Allowing the inner slope of the satellite LRG density profile to become steeper than $r^{-1}$ is exactly what was needed to match the data.  The value of $\gamma$ is well constrained and our MCMC yields $\gamma= 2.06 \pm 0.21$.  It is certainly not surprising that the inner slope of the satellite density profile is similar to the slope of $\xi(r)$ at small scales because, as we argued in \S~\ref{2pt_xigg}, most LRG pairs should be central-satellite pairs whose pair distribution $F_{\mathrm{cs}}(r)$ is essentially the density profile itself.

We have established that neither $P(N|M)$, nor the concentration of satellite LRGs, are sufficient to explain the M06 small-scale data, and that a profile other than NFW is needed.  We thus now allow our model to have more than 4 free parameters and we vary both $\gamma$ and $\fgal$.  Since our density profile is no longer NFW, there is no reason to keep the concentration fixed to what was found for NFW dark matter halos.  In our final model, we thus vary 5 parameters: $\Mmin$, $\Mone$, $\alpha$, $\fgal$, and $\gamma$, and we refer to this model as PNMCG.  Our goal for investigating this model is to determine exactly what constraints the M06 data place on the density profile of satellite LRGs.

\begin{figure}[t]
\begin{center}
\includegraphics[scale=0.55,angle=0]{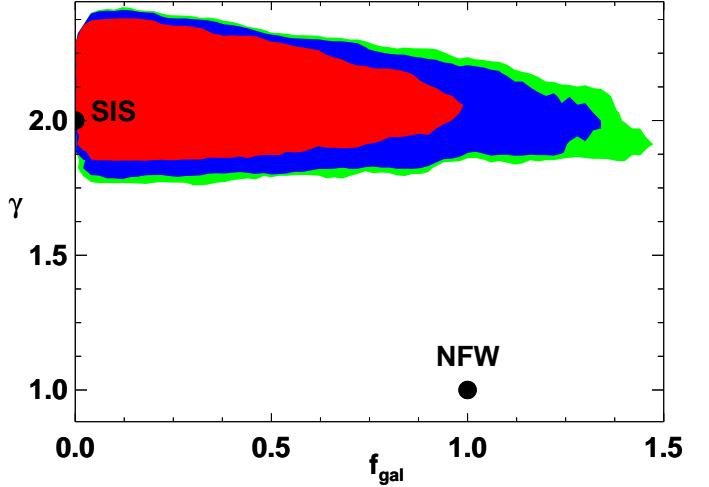}
\caption{Allowed density profiles for LRG satellite galaxies within their dark matter halos.  Shaded regions show the 1-$\sigma$ ({\it red}), 2-$\sigma$ ({\it blue}) and 3-$\sigma$ ({\it green}) allowed regions for $\gamma$ and $\fgal$ for the PNMCG model.  Filled circles mark the locations of the NFW profile ($\gamma =1$, $\fgal =1$) and the Singular Isothermal Sphere (SIS) profile ($\gamma = 2$, $\fgal \rightarrow 0$).}
\label{fig:hist_Slope}
\end{center}
\end{figure}

We find a best-fit model with a reduced $\chi^2$ of 0.71 ($\chi^2$ of 2.82 with 4 degrees of freedom).  Figure~\ref{fig:hist_Slope} shows our 1-,2- and 3-$\sigma$ contours for $\gamma$ and $\fgal$.  As before, we find that the satellite LRG profile is much steeper than NFW (for NFW, $\gamma =1$ and $\fgal =1$), with $\gamma = -2.17\pm 0.12$.  As we argued in \S~\ref{HOD}, most of the LRG satellites reside in halos of mass close to $\Mone$.  At our best-fit value for $\Mone$ ($10^{14.62}\hMsun$), the dark matter concentration fitting formula from \S~\ref{HOD} gives a halo concentration of $c = 4.25$.  Applying our 1-$\sigma$ range for $\fgal$ implies that $\Cgal\sim 0.1 - 4.1$.  Concentration values of unity or less mean that the scale radius $r_s$ is larger than the virial radius, which essentially means that the density profile retains its inner slope most of the way out.  In other words, our fit shows that the density profile for LRG satellites is consistent with a simple isothermal profile.

\section{DISCUSSION}\label{discussion}

Our results show that the distribution of satellite LRGs within dark matter halos requires a steeper inner density profile than NFW, which suggests that these galaxies are poor tracers of the dark matter distribution at these scales.  The density profile of dark matter halos in the $\Lambda$CDM model has been measured extensively using high resolution N-body simulations, and recent inner profile measurements seem to confirm the NFW $\sim r^{-1}$ predictions \citep{navarro08} -- with some slight deviations found in separate work \citep{diemand04,fukushige04,reed05,delpopolo09}.  However, NFW does not consider baryons, which can affect the dark matter density profile at small scales.  The interaction between baryons and dark matter is addressed by the adiabatic contraction model that describes the gravitational effect of baryons on dark matter as the gas condenses and sinks to the center of the dark matter potential well.  The gravitational influence of the baryons draws the dark matter in, and this can steepen the density profile \citep{gnedin04, romanodiaz08, weinberg08, sommerlarsen09}.  Although it has been shown that the inner profile can significantly steepen \citep{gustafsson06}, the majority of results show only a moderate steepening.  This has also been observationally confirmed using galaxy-galaxy lensing by \citet{mandelbaum06a} who find that the mass density profile of LRG clusters is consistent with NFW.  LRG satellites therefore have a steeper density profile than dark matter even with the effects of baryons taken into consideration.

It is not necessarily surprising that LRGs are poor tracers of the dark matter density distribution within halos.  LRGs presumably live in subhalos, which can certainly have a different distribution than their host halos.  \citet{nagai05} found that subhalos actually have a shallower profile than dark matter at larger scales, but this has not been studied for the massive halos and small scales we consider here.  It is difficult to model this regime because simulations must have, both a very large volume that contains many cluster-sized halos, and high mass and spatial resolution to resolve scales of $\sim10\hkpc$.  In fact, on these very small scales, it is likely that dark matter subhalos have already been disrupted by tidal forces, while the LRGs, being smaller and denser, have survived.  So pure dark matter simulations may be insufficient to predict these LRG results.  However, semi-analytic models that include galaxies and can have arbitrarily high resolution should be able to make these predictions (see \citealt{kitzbichler08} for semi-analytic modeling of the very-small scale clustering of lower luminosity galaxies).

In any case, our results have implications for the modeling of LRG clustering because a standard NFW profile cannot describe the spatial distribution of LRGs on scales $\lesssim 0.03\hmpc$.  It would be interesting to see if lower luminosity galaxies exhibit the same behavior or if this is simply a feature for LRGs.  It would also be interesting to see if LRGs maintain their steep density profile at high redshift.


\section{ACKNOWLEDGEMENTS}\label{acknowledgements}

We thank Joanna Dunkley and David Hogg for useful discussions and comments.  We thank Zheng Zheng for providing us with his best-fit 2-halo term for LRGs.


\begin{thebibliography}{37}
\expandafter\ifx\csname natexlab\endcsname\relax\def\natexlab#1{#1}\fi

\bibitem[{{Berlind} \& {Weinberg}(2002)}]{berlind02a}
{Berlind}, A.~A., \& {Weinberg}, D.~H. 2002, \apj, 575, 587

\bibitem[{{Bullock} {et~al.}(2001){Bullock}, {Kolatt}, {Sigad}, {Somerville},
  {Kravtsov}, {Klypin}, {Primack}, \& {Dekel}}]{bullock01}
{Bullock}, J.~S., {Kolatt}, T.~S., {Sigad}, Y., {Somerville}, R.~S.,
  {Kravtsov}, A.~V., {Klypin}, A.~A., {Primack}, J.~R., \& {Dekel}, A. 2001,
  \mnras, 321, 559

\bibitem[{{Cooray} \& {Sheth}(2002)}]{cooray02}
{Cooray}, A., \& {Sheth}, R. 2002, \physrep, 372, 1

\bibitem[{{Davis} \& {Peebles}(1983)}]{davis83}
{Davis}, M., \& {Peebles}, P.~J.~E. 1983, \apj, 267, 465

\bibitem[{{Del Popolo} \& {Kroupa}(2009)}]{delpopolo09}
{Del Popolo}, A., \& {Kroupa}, P. 2009, ArXiv e-prints

\bibitem[{{Diemand} {et~al.}(2004){Diemand}, {Moore}, \& {Stadel}}]{diemand04}
{Diemand}, J., {Moore}, B., \& {Stadel}, J. 2004, \mnras, 353, 624

\bibitem[{{Dunkley} {et~al.}(2005){Dunkley}, {Bucher}, {Ferreira}, {Moodley},
  \& {Skordis}}]{dunkley05}
{Dunkley}, J., {Bucher}, M., {Ferreira}, P.~G., {Moodley}, K., \& {Skordis}, C.
  2005, \mnras, 356, 925

\bibitem[{{Eisenstein} {et~al.}(2005){Eisenstein}, {Zehavi}, {Hogg},
  {Scoccimarro}, {Blanton}, {Nichol}, {Scranton}, {Seo}, {Tegmark}, {Zheng},
  {Anderson}, {Annis}, {Bahcall}, {Brinkmann}, {Burles}, {Castander},
  {Connolly}, {Csabai}, {Doi}, {Fukugita}, {Frieman}, {Glazebrook}, {Gunn},
  {Hendry}, {Hennessy}, {Ivezi{\'c}}, {Kent}, {Knapp}, {Lin}, {Loh}, {Lupton},
  {Margon}, {McKay}, {Meiksin}, {Munn}, {Pope}, {Richmond}, {Schlegel},
  {Schneider}, {Shimasaku}, {Stoughton}, {Strauss}, {SubbaRao}, {Szalay},
  {Szapudi}, {Tucker}, {Yanny}, \& {York}}]{eisenstein05b}
{Eisenstein}, D.~J., {et~al.} 2005, \apj, 633, 560

\bibitem[{{Fukushige} {et~al.}(2004){Fukushige}, {Kawai}, \&
  {Makino}}]{fukushige04}
{Fukushige}, T., {Kawai}, A., \& {Makino}, J. 2004, \apj, 606, 625

\bibitem[{{Gnedin} {et~al.}(2004){Gnedin}, {Kravtsov}, {Klypin}, \&
  {Nagai}}]{gnedin04}
{Gnedin}, O.~Y., {Kravtsov}, A.~V., {Klypin}, A.~A., \& {Nagai}, D. 2004, \apj,
  616, 16

\bibitem[{{Gustafsson} {et~al.}(2006){Gustafsson}, {Fairbairn}, \&
  {Sommer-Larsen}}]{gustafsson06}
{Gustafsson}, M., {Fairbairn}, M., \& {Sommer-Larsen}, J. 2006, \prd, 74,
  123522

\bibitem[{{Kitzbichler} \& {White}(2008)}]{kitzbichler08}
{Kitzbichler}, M.~G., \& {White}, S.~D.~M. 2008, \mnras, 391, 1489

\bibitem[{{Kravtsov} {et~al.}(2004){Kravtsov}, {Berlind}, {Wechsler}, {Klypin},
  {Gottl{\"o}ber}, {Allgood}, \& {Primack}}]{kravtsov04b}
{Kravtsov}, A.~V., {Berlind}, A.~A., {Wechsler}, R.~H., {Klypin}, A.~A.,
  {Gottl{\"o}ber}, S., {Allgood}, B., \& {Primack}, J.~R. 2004, \apj, 609, 35

\bibitem[{{Li} \& {White}(2009)}]{li09}
{Li}, C., \& {White}, S.~D.~M. 2009, ArXiv e-prints

\bibitem[{{Lupton} {et~al.}(2001){Lupton}, {Gunn}, {Ivezi{\'c}}, {Knapp}, \&
  {Kent}}]{lupton01}
{Lupton}, R., {Gunn}, J.~E., {Ivezi{\'c}}, Z., {Knapp}, G.~R., \& {Kent}, S.
  2001, in Astronomical Society of the Pacific Conference Series, Vol. 238,
  Astronomical Data Analysis Software and Systems X, ed. F.~R. {Harnden}, Jr.,
  F.~A. {Primini}, \& H.~E. {Payne}, 269--+

\bibitem[{{Mandelbaum} {et~al.}(2006){Mandelbaum}, {Seljak}, {Cool}, {Blanton},
  {Hirata}, \& {Brinkmann}}]{mandelbaum06a}
{Mandelbaum}, R., {Seljak}, U., {Cool}, R.~J., {Blanton}, M., {Hirata}, C.~M.,
  \& {Brinkmann}, J. 2006, \mnras, 372, 758

\bibitem[{{Masjedi} {et~al.}(2006){Masjedi}, {Hogg}, {Cool}, {Eisenstein},
  {Blanton}, {Zehavi}, {Berlind}, {Bell}, {Schneider}, {Warren}, \&
  {Brinkmann}}]{masjedi06a}
{Masjedi}, M., {et~al.} 2006, \apj, 644, 54

\bibitem[{{Nagai} \& {Kravtsov}(2005)}]{nagai05}
{Nagai}, D., \& {Kravtsov}, A.~V. 2005, \apj, 618, 557

\bibitem[{{Navarro} {et~al.}(1997){Navarro}, {Frenk}, \& {White}}]{nfw97}
{Navarro}, J.~F., {Frenk}, C.~S., \& {White}, S.~D.~M. 1997, \apj, 490, 493

\bibitem[{{Navarro} {et~al.}(2008){Navarro}, {Ludlow}, {Springel}, {Wang},
  {Vogelsberger}, {White}, {Jenkins}, {Frenk}, \& {Helmi}}]{navarro08}
{Navarro}, J.~F., {et~al.} 2008, ArXiv e-prints

\bibitem[{{Peacock} \& {Smith}(2000)}]{peacock00a}
{Peacock}, J.~A., \& {Smith}, R.~E. 2000, \mnras, 318, 1144

\bibitem[{{Reed} {et~al.}(2005){Reed}, {Governato}, {Verde}, {Gardner},
  {Quinn}, {Stadel}, {Merritt}, \& {Lake}}]{reed05}
{Reed}, D., {Governato}, F., {Verde}, L., {Gardner}, J., {Quinn}, T., {Stadel},
  J., {Merritt}, D., \& {Lake}, G. 2005, \mnras, 357, 82

\bibitem[{{Romano-D{\'{\i}}az} {et~al.}(2008){Romano-D{\'{\i}}az}, {Shlosman},
  {Hoffman}, \& {Heller}}]{romanodiaz08}
{Romano-D{\'{\i}}az}, E., {Shlosman}, I., {Hoffman}, Y., \& {Heller}, C. 2008,
  \apjl, 685, L105

\bibitem[{{Scoccimarro} {et~al.}(2001){Scoccimarro}, {Sheth}, {Hui}, \&
  {Jain}}]{scoccimarro01a}
{Scoccimarro}, R., {Sheth}, R.~K., {Hui}, L., \& {Jain}, B. 2001, \apj, 546, 20

\bibitem[{{Sheth} {et~al.}(2001){Sheth}, {Diaferio}, {Hui}, \&
  {Scoccimarro}}]{sheth01}
{Sheth}, R.~K., {Diaferio}, A., {Hui}, L., \& {Scoccimarro}, R. 2001, \mnras,
  326, 463

\bibitem[{{Sommer-Larsen} \& {Limousin}(2009)}]{sommerlarsen09}
{Sommer-Larsen}, J., \& {Limousin}, M. 2009, ArXiv e-prints

\bibitem[{{Tinker}(2007)}]{tinker07}
{Tinker}, J.~L. 2007, \mnras, 374, 477

\bibitem[{{Wang} {et~al.}(2006){Wang}, {Li}, {Kauffmann}, \& {De
  Lucia}}]{wang06}
{Wang}, L., {Li}, C., {Kauffmann}, G., \& {De Lucia}, G. 2006, \mnras, 371, 537

\bibitem[{{Warren} {et~al.}(2006){Warren}, {Abazajian}, {Holz}, \&
  {Teodoro}}]{warren06}
{Warren}, M.~S., {Abazajian}, K., {Holz}, D.~E., \& {Teodoro}, L. 2006, \apj,
  646, 881

\bibitem[{{Weinberg} {et~al.}(2008){Weinberg}, {Colombi}, {Dav{\'e}}, \&
  {Katz}}]{weinberg08}
{Weinberg}, D.~H., {Colombi}, S., {Dav{\'e}}, R., \& {Katz}, N. 2008, \apj,
  678, 6

\bibitem[{{York} {et~al.}(2000){York}, {Adelman}, {Anderson}, {Anderson},
  {Annis}, {Bahcall}, {Bakken}, {Barkhouser}, {Bastian}, {Berman}, {Boroski},
  {Bracker}, {Briegel}, {Briggs}, {Brinkmann}, {Brunner}, {Burles}, {Carey},
  {Carr}, {Castander}, {Chen}, {Colestock}, {Connolly}, {Crocker}, {Csabai},
  {Czarapata}, {Davis}, {Doi}, {Dombeck}, {Eisenstein}, {Ellman}, {Elms},
  {Evans}, {Fan}, {Federwitz}, {Fiscelli}, {Friedman}, {Frieman}, {Fukugita},
  {Gillespie}, {Gunn}, {Gurbani}, {de Haas}, {Haldeman}, {Harris}, {Hayes},
  {Heckman}, {Hennessy}, {Hindsley}, {Holm}, {Holmgren}, {Huang}, {Hull},
  {Husby}, {Ichikawa}, {Ichikawa}, {Ivezi{\'c}}, {Kent}, {Kim}, {Kinney},
  {Klaene}, {Kleinman}, {Kleinman}, {Knapp}, {Korienek}, {Kron}, {Kunszt},
  {Lamb}, {Lee}, {Leger}, {Limmongkol}, {Lindenmeyer}, {Long}, {Loomis},
  {Loveday}, {Lucinio}, {Lupton}, {MacKinnon}, {Mannery}, {Mantsch}, {Margon},
  {McGehee}, {McKay}, {Meiksin}, {Merelli}, {Monet}, {Munn}, {Narayanan},
  {Nash}, {Neilsen}, {Neswold}, {Newberg}, {Nichol}, {Nicinski}, {Nonino},
  {Okada}, {Okamura}, {Ostriker}, {Owen}, {Pauls}, {Peoples}, {Peterson},
  {Petravick}, {Pier}, {Pope}, {Pordes}, {Prosapio}, {Rechenmacher}, {Quinn},
  {Richards}, {Richmond}, {Rivetta}, {Rockosi}, {Ruthmansdorfer}, {Sandford},
  {Schlegel}, {Schneider}, {Sekiguchi}, {Sergey}, {Shimasaku}, {Siegmund},
  {Smee}, {Smith}, {Snedden}, {Stone}, {Stoughton}, {Strauss}, {Stubbs},
  {SubbaRao}, {Szalay}, {Szapudi}, {Szokoly}, {Thakar}, {Tremonti}, {Tucker},
  {Uomoto}, {Vanden Berk}, {Vogeley}, {Waddell}, {Wang}, {Watanabe},
  {Weinberg}, {Yanny}, \& {Yasuda}}]{york00a}
{York}, D.~G., {et~al.} 2000, \aj, 120, 1579

\bibitem[{{Zehavi} {et~al.}(2005){Zehavi}, {Eisenstein}, {Nichol}, {Blanton},
  {Hogg}, {Brinkmann}, {Loveday}, {Meiksin}, {Schneider}, \&
  {Tegmark}}]{zehavi05b}
{Zehavi}, I., {et~al.} 2005, \apj, 621, 22

\bibitem[{{Zehavi} {et~al.}(2004){Zehavi}, {Weinberg}, {Zheng}, {Berlind},
  {Frieman}, {Scoccimarro}, {Sheth}, {Blanton}, {Tegmark}, {Mo}, {Bahcall},
  {Brinkmann}, {Burles}, {Csabai}, {Fukugita}, {Gunn}, {Lamb}, {Loveday},
  {Lupton}, {Meiksin}, {Munn}, {Nichol}, {Schlegel}, {Schneider}, {SubbaRao},
  {Szalay}, {Uomoto}, \& {York}}]{zehavi04a}
------. 2004, \apj, 608, 16

\bibitem[{{Zheng}(2004)}]{zheng04a}
{Zheng}, Z. 2004, \apj, 610, 61

\bibitem[{{Zheng} {et~al.}(2005){Zheng}, {Berlind}, {Weinberg}, {Benson},
  {Baugh}, {Cole}, {Dav{\'e}}, {Frenk}, {Katz}, \& {Lacey}}]{zheng05}
{Zheng}, Z., {et~al.} 2005, \apj, 633, 791

\bibitem[{{Zheng} {et~al.}(2007){Zheng}, {Coil}, \& {Zehavi}}]{zheng07}
{Zheng}, Z., {Coil}, A.~L., \& {Zehavi}, I. 2007, \apj, 667, 760

\bibitem[{{Zheng} {et~al.}(2008){Zheng}, {Zehavi}, {Eisenstein}, {Weinberg}, \&
  {Jing}}]{zheng08}
{Zheng}, Z., {Zehavi}, I., {Eisenstein}, D.~J., {Weinberg}, D.~H., \& {Jing},
  Y. 2008, ArXiv e-prints

\end{thebibliography}
\end{document}